\documentclass[useAMS]{mn2e}
\usepackage{graphicx}
\usepackage{times}

\usepackage{colordvi}
\usepackage{epsfig}
\usepackage{appendix}
\title[Dust in high redshift DLAs]{ Exploring the dust content of SDSS DR7
damped Lyman alpha systems at 2.15$\le z_{ab}< $5.2} \author[P.  Khare et al.]
{Pushpa Khare$^1$\thanks{E-mail: pushpakhare@gmail.com}, Daniel Vanden
Berk$^2$, Donald G.  York$^{3,4}$, Britt Lundgren$^{5}$ , Varsha P.
Kulkarni$^6$\\ $^1$CSIR Emeritus Scientist, IUCAA, Ganeshkhind, Pune, 411007,
India\\$^2$Physics Department, St. Vincent College, Latrobe, PA 15650, USA \\
$^3$Department of Astronomy and Astrophysics, University of Chicago, Chicago,
IL 60637, USA \\ $^4$Enrico Fermi Institute, University of Chicago, Chicago, IL
60637, USA \\ $^5$ Department of Astronomy, Yale University, New Haven, CT
06520-8101, USA\\ $^6$ Department of Physics and Astronomy, University of South
Carolina, Columbia, SC 29208, USA\\}
\begin{document}
\date{}
\maketitle
\label{firstpage}
\begin{abstract}
We have studied a sample of 1084 intervening absorption systems with 2.15$\le
z_{ab}\le$5.2, having log(N$_{\rm HI}$) $>$ 20.0 in the spectra of QSOs in
Sloan Digital Sky Survey (SDSS) data release 7 (DR7), with the aim of
understanding the nature and abundance of the dust and the chemical abundances
in the DLA absorbers. Composite spectra were constructed for the full sample
and several subsamples, chosen on the basis of absorber and QSO properties.
Average extinction curves were obtained for the samples by comparing their
geometric mean composite spectra with those of two samples of QSOs, matching in
z$_{em}$ and i magnitude with the DLA sample, one sample without any absorbers
along their lines of sight and the other without any DLAs along their lines of
sight irrespective of the presence of other absorption systems. We also derived
relative extinction curves of several pairs of subsamples. While the average
reddening in the DLA sample is small, we find definite evidence for the
presence of dust in subsamples based on absorber properties, in particular the
strength of metal absorption lines. DLAs along lines of sight to QSOs which are
not colour selected are found to be more dusty compared to those along the
lines of sight to the more numerous colour selected QSOs. From these studies
and from the strengths of absorption lines in the composite spectra, we
conclude that $\le$ 10\% of the DLAs in SDSS DR7 cause significant reddening,
have stronger absorption lines and have higher abundances as compared to the
rest of the sample. The rest of the sample shows little reddening.  While due
to the dominant color selection method used to target QSOs in the SDSS DR7,
this fraction of 10\% likely represents a lower limit for the global fraction
of dusty DLAs at high-z, it is also possible that the dust grain sizes at high
redshifts are larger, giving rise to a flat extinction curve over the observed
range of wavelengths.
  
\end{abstract}
\section {Introduction}
Early attempts to quantitatively detect dust in damped Lyman alpha absorbers in
QSO spectra, by its differential extinction of a given object, were made by
Pei, Fall \& Bechtold (1991). They found higher extinction in a sample of 13
QSOs having damped Ly\, $\alpha$ systems (DLAs) in their spectra compared to a
sample of 15 QSOs that had no DLAs indicating presence of dust in the DLA
absorbers, albeit with $E(B-V) < $0.03. Ellison, Hall \& Lira (2005) similarly
found only a slight reddening ($E(B-V) <$ 0.04) in 14 out of 42 QSOs with DLAs.

The Sloan Digital Sky Survey (SDSS) archive has been used for several studies
of extinction in sight-lines to QSOs. Richards et al. (2003) showed that an
observer frame colour excess, hereafter, $\Delta(g-i)$, which is the difference
between the actual colours of a QSO and the median colours of QSOs at that
redshift, could be defined for the SDSS QSO spectra and could be used to form
templates of objects with apparent degrees of extinction. However, they could
not discern if the extinction so detected was from the QSO itself or from
intervening systems.  Hopkins et al.  (2004) used composite QSO spectra to show
that the extinction towards QSOs is dominated by SMC-like extinction, which
they argued was predominantly located at QSO redshifts. Wild \& Hewett (2005)
have found evidence of dust in QSO absorption line systems with detected Ca II
lines, with $E(B-V)$ of 0.06. Wild, Hewitt \& Pettini (2005) find strong Ca II
absorbers to have $E(B-V)>0.1$.  However, these samples are small and highly
selective.  Nestor et  al. (2008) find no evidence that N$_{\rm Ca II}$ in DLAs
is related to the presence of dust depletion of Cr. Murphy \& Liske (2004)
studied the spectral indices of QSOs in SDSS Data Release 2 and found no
evidence for the presence of dust in DLAs at a redshift $\sim$ 3. They derived
an upper limit of 0.02 on $E(B-V)$.

York et al. (2006; hereafter Y06) gave definite evidence of dust in the
intervening Mg II absorbers by comparing composite spectra of SDSS QSOs with
absorbers with that of a matching (in z$_{em}$ and i magnitude (m$_i$)) sample
(the non-absorber sample) of SDSS QSOs without any absorbers in their spectra.
The reddening was shown to be correlated with several absorber properties, in
particular the strength of Mg II lines (essentially, the velocity spread of
multiple, saturated components).  Vanden Berk et al. (2008), using the method
of Y06, similarly obtained evidence for reddening in systems associated with
QSOs (having relative velocity with respect to the QSO smaller than 3000 km
s$^{-1}$). M{\'e}nard et al. (2008), investigated the effect of dust reddening
and gravitational lensing of Mg II absorbers in SDSS DR4 QSOs and derived the
dependence of reddening on redshift and on Mg II equivalent width.

Recently, Frank and P{\'e}roux (2010; hereafter FP10) used the method of Y06 to
study the dust content in intervening DLAs at redshifts of $>$ 2.2. They found
no evidence of dust in their sample of 731 DLAs and subsamples there of. This
result is surprising in view of the fact that several of the high N$_{\rm HI}$
systems at 1.8$<$z$_{ab}<4.2$ have been found to have molecular hydrogen which
appears to be associated with significant depletion of Fe (Noterdaeme et al.
2008). Molecular hydrogen is believed to form on the surface of dust grains
(Hollenbach \& Salpeter 1970) and depletion of Fe is thought to be related, in
some cases, to gas phase depletion onto solid particles.  The result of FP10
thus prompted us to further investigate the dust content of DLAs.  We do note
that Tumlinson et al. (2002) have found modest amounts of H$_2$ in the LMC and
the SMC along sight lines having low $E(B-V)$. Both these clouds are known to
have dust properties different from those of the dust in the Milky Way.

In this paper, we present the results of our investigations. Section 2
describes our sample selection and generation of composite spectra, section 3
describes the results and conclusions are presented in section 4.

\section{Sample selection and generation of composite spectra} A sample of
intervening absorption systems (with relative velocities with respect to QSOs
$>$ 5000 km s$^{-1}$) having neutral hydrogen column densities $\ge10^{20}$
cm$^{-2}$ in the SDSS DR7 has been compiled by Noterdaeme et al. (2009b). We
used their sample which consists of 1426 systems at redshifts between 2.15 and
5.2.  

Following FP10, we removed QSOs having more than one of these systems along the
line of sight. Quasars containing likely broad absorption line systems
(BALs) were removed from the sample and from the comparison sample of matched
quasars (described below). All BALQSOs identified in the SDSS Data Release 5
(DR5) catalog by Gibson et al. (2009) were removed. We manually searched for
BALQSOs among all of the quasars in the DR7 quasar catalog (Schneider et al.
2010) that are not also in the DR5 catalog.  The number of quasars inspected is
28532.  A broad absorption line system candidate was defined as any absorption
feature that appeared to be at least broad enough that it could be a C\,{\sc
iv}, Mg\,{\sc ii}, or some other doublet line with heavily blended components.
That is, any feature broad enough that a common absorption doublet could not be
visually resolved in the spectrum was considered a candidate BAL.  (We call the
objects ``candidate'' BALs because they have not been quantitatively analyzed
to determine if they meet various criteria listed in published BALQSO
catalogs.) The selection procedure using this definition was simple to
implement, and the results corresponded well with those in the Gibson et al.
(2009) catalog.  We also inspected all of the post DR5 candidate BALQSOs
identified by Shen et al. (2011), using the same selection
criteria.  Each post DR5 BALQSO candidate was inspected by at least three
people.  A total of 2206 BALQSO candidates were identified, among 28532
quasars.  The candidate BALQSO fraction ($7.7\%$) in the post DR5 quasar sample
is slightly larger than the BALQSO fraction ($6.5\%$) in the Gibson et al.
(2009) catalog.  Thus we have likely selected somewhat more candidate BALQSOs
than the Gibson et. al.  procedure would have.  For the purposes of this paper,
the difference is minor, and it means that we are slightly less likely to
include quasars with strong (non-DLA) absorption among the matched samples.

This left us with 1084 systems out of which 721 systems are classical DLAs,
having log(N$_{\rm HI}$) $\ge$ 20.3 and the rest, 363 are sub-DLAs, having
log(N$_{\rm HI}$) between 20 and 20.3.  These 1084 systems form our primary
absorber sample S1. The redshift distribution of S1 is shown in Fig.1. It
can be seen that most (93\%) of the systems in S1 have redshifts between 2.25
and 4. Several subsamples constructed from S1 are listed in Table 1 which
lists the selection criteria, number of systems in the samples, and average
values of N$_{\rm HI}$, z$_{ab}$ and m$_i$ for the samples. Values of $E(B-V)$
(described below) are also given.

We also inspected the matched non-DLA sample of FP10 (matched spectrum data
list sent courtesy of Stephan Frank) to check whether any BALQSOs may have been
included in that sample.  Of the 731 matched quasars used by FP10, there are 60
candidate BALQSOs from our combined SDSS DR7 list. In addition there are 29
DLAs which have inexplicably found their way into the matching non-DLA sample
used by FP10, most of which are matches with the DLA itself. The inclusion of
BALQSOs and DLA quasars in the matched quasar sample {\em may} explain at least
in part why FP10 did not detect significant reddening associated with DLAs
---the matched sample may have been artificially reddened due to the inclusion
of BALQSOs which are known to be dusty and quasars with DLAs. We have
therefore, repeated some of the analysis of FP10.

Below we briefly describe the method of generation and use of composite
spectra as advocated by Y06. Further details can be found in Y06. To generate
an arithmetic mean composite, the spectra were first normalized by
reconstructions of the QSO continua, using the first 30 QSO eigenspectra
derived by Yip et al. (2004).  The spectra were then shifted to the absorber
rest frame. The pixels flagged by the spectroscopic pipeline as possibly bad in
some way (Stoughton et al. 2002) were masked and not used in constructing the
composites. Also masked were pixels within 5 {\AA} of the expected line
positions of detected absorption systems unrelated to the target system. After
masking pixels, the normalized flux density in each remaining pixel was
weighted by the inverse of the associated variance, and the weighted arithmetic
mean of all contributing spectra was calculated for each pixel.  

\begin{table*}
\centering
\begin{minipage}{140mm}
\caption{Samples: Definitions and properties}
\small
\begin{tabular}
{|l|l|r|r|r|r|r|r|r|}
\hline
Sample&Selection
Criteria&Number&$E(B-V)^a$&$E(B-V)^b$&$<z_{ab}>$
&log$<$N$_{\rm H I}>$&$<$m$_i>$&$<\Delta(g-i)>$ \\
&&of systems&&&&&&\\
\hline
S1& Full sample & 1084&0.0014&-0.0026&2.9323&20.72&18.98&-0.011\\
S2& m$^c_i<$19.5 & 847&0.0046&0.0005&2.8914&20.74&18.77&-0.014\\
S3&m$_i<$19.5,&545&0.0044$^d$&0.0008$^d$&2.8436&20.72&18.99&0.000\\
&$|\Delta$ M$_i^e|^f<0.1$ \&&799$^g$&&&2.8703&20.74&18.82&-0.001\\
& $|\Delta$z$_{em}|^f < 0.05$&&&&&&&\\
S4&m$_i<$19.5, &609&0.0050&0.0007&2.8644&20.73&18.96&-0.001\\
&$|\Delta$ M$_i^e|<0.15$ \&&830$^g$&&&2.8644&20.75&18.80&0.000\\
& $|\Delta$z$_{em}| < 0.1$&&&&&&&\\
S5&log(N$_{\rm HI}) \ge 20.49$&538&0.0028&-0.0009&2.9484&20.95&18.95&-0.008\\
S6&log(N$_{\rm HI}) < 20.49$&546&0.0000&-0.0043&2.9164&20.25&19.01&-0.014\\
S7&m$_i\ge$19.05&545&-0.0036&-0.0072&3.0814&20.69&19.47&0.002\\
S8&m$_i<$19.05&539&0.0053&0.0008&2.7815&20.76&18.49&-0.024\\
S9&M$_i\ge$-27.42&546&-0.0022&-0.0068&2.8614&20.68&19.39&0.00\\
S10&M$_i<$-27.42&538&0.0065&0.0027&3.0042&20.76&18.57&-0.022\\
S11&z$_{ab}\ge2.89$&543&-0.0008&-0.0048&3.2943&20.75&19.14&0.014\\
S12&z$_{ab}<2.89$&541&0.0025&-0.0013&2.5688&20.69&18.83&-0.036\\
S13&z$_{em}\ge3.26$&538&-0.0001&-0.0036&3.2370&20.76&19.10&0.033\\
S14&z$_{em}<3.26$&546&0.0019&-0.0020&2.6320&20.69&18.86&-0.054\\
S15&$\Delta(g-i)\ge-0.01$&547&0.0110&0.0068&2.9254&20.79&18.91&-0.012\\
S16&$\Delta(g-i)<-0.01$&537&-0.0090&-0.0131&2.9393&20.64&19.06&-0.009\\
S17&W$_{\rm SiII}^h\ge0.62$&213&0.0058&0.0010&2.8740&20.93&19.02&-0.015\\
S18&W$_{\rm SiII}<0.62$&226&0.0045&-0.0022&2.7281&20.73&18.48&-0.062\\
S19&log(N$_{\rm HI})\ge21.0$&149&0.0081&-0.0014&2.9903&21.26&18.89&0.008\\
S20&log(N$_{\rm HI})<21.0$&935&0.0004&-0.0030&2.9230&20.51&19.00&-0.014\\
S21&$\Delta(g-i)\ge0.3$&60&0.0290&0.0290&3.1925&20.93&18.93&0.007\\
S22&$\Delta(g-i)<0.3$&1024&-0.0002&-0.0045&2.9170&20.71&18.99&-0.012\\
S23&W$_{\rm SiII}\ge1.0$&106&0.0097&0.0054&2.9526&20.99&19.09&-0.009\\
S24&W$_{\rm SiII}<1.0$ &333&0.0039&-0.0026&2.7500&20.78&18.64&-0.049\\
S25&W$_{\rm SiII}\ge1.5$&56&0.0092&-0.0008&3.0427&21.04&19.04&-0.001\\
S26&W$_{\rm SiII}<1.5$&383&0.0049&-0.0008&2.7633&20.80&18.70&-0.045\\
\hline
\multicolumn{7}{l}{ a: Extinction of the absorber sample with respect to the
corresponding non-absorber sample assuming SMC extinction law}\\ 
\multicolumn{7}{l}{ b: Extinction of the absorber sample with respect to the
corresponding non-DLA sample assuming SMC extinction law}\\ 
\multicolumn{7}{l}{ c: i magnitude corrected for Galactic extinction}\\
\multicolumn{7}{l}{ d: One $\sigma$ error from bootstrap analysis is $\simeq$0.0015}\\
\multicolumn{7}{l}{ e: Absolute i magnitude}\\
\multicolumn{7}{l}{ f: Absolute values of the difference between absorber and non-absorber QSO pairs}\\
\multicolumn{7}{l}{ g: Values in this row are obtained when non-DLA sample is used as the matching sample.}\\
\multicolumn{7}{l}{ h: Rest equivalent width of Si II$\lambda$1526}\\
\end{tabular}
\end{minipage}
\end{table*}

The geometric mean of a set of power law spectra preserves the average power
law index of the spectra, so it is the appropriate statistic to use to
determine the characteristic extinction law (likely to be approximated by a
power law) of the absorber sample. The geometric mean composite spectrum for an
absorber sample was generated by shifting the spectra to the absorber rest
frames and taking their geometric mean. The non-absorber sample was compiled by
selecting one non-absorber QSO (not having any absorption lines in its
spectrum) for each QSO in the absorber sample, having nearly the same  z$_{em}$
and M$_i$ (see Y06 for details). To generate the geometric mean of the
non-absorber sample, each non-absorber spectrum was first shifted to the rest
frame of the absorber in the corresponding absorber QSO. As long as the
spectroscopic properties of the QSOs in the absorber and non-absorber samples
are statistically equivalent (which is ensured by their having similar z$_{em}$
and M$_i$), except for the effects of the intervening absorption systems,
dividing the absorber composite by the non-absorber composite will yield the
relative extinction curve in the absorber frame. In doing this, the
non-absorber spectrum is scaled so that the average flux densities in the
reddest 100 {\AA} of both spectra are equal.  The extinction curve can be
fitted with the SMC, LMC or Milky Way (MW) extinction curves to obtain the
average $E(B-V)$ for the absorber sample. For generating both types of
composites, we used only parts of spectra having $\lambda > 1250$ {\AA} in the
rest frame of QSOs in order to avoid any contamination by the Lyman alpha
forest in the spectra.

Although FP10 use the method of composite spectra as advocated by Y06, their
procedure differed from Y06 in respect of the use of the matching non-absorber
sample.  They used the sample of QSOs not having DLAs (non-DLA sample) as their
non-absorber sample, arguing that any other (lower H I column density) systems
will be found with equal probability in the absorber as well as the non-DLA
sample and the effect of these systems should cancel out.

\begin{figure}
\includegraphics[width=3.15in,height=3.0in]{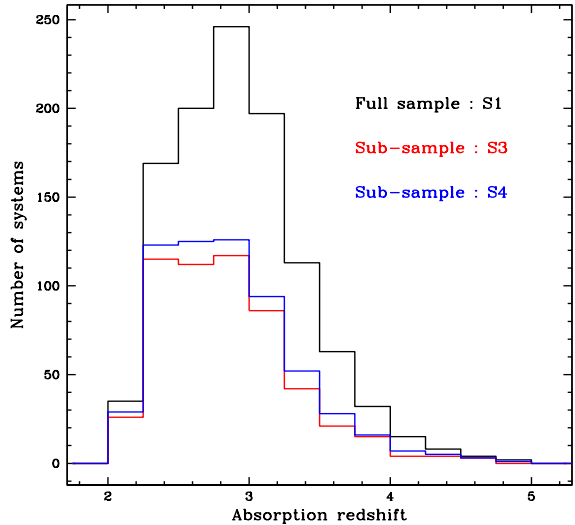}
\caption{Histograms showing the redshift distribution of systems in samples S1,
S3 and S4. About 93-94\% of the systems in all three samples have redshifts
between 2.25 and 4.}
\end{figure}

\begin{figure}
\includegraphics[width=3.15in,height=3.0in]{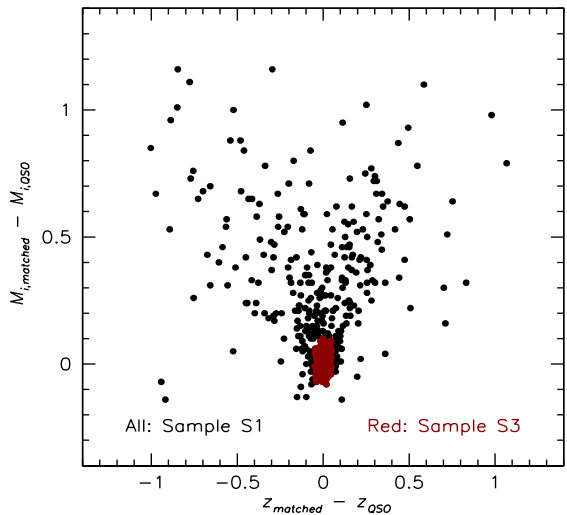}
\caption{Plot showing the differences between z$_{em}$ and M$_i$ of the QSOs in
the full DLA sample (S1) and the matching non-absorber sample. In some cases
the differences are large, with $|\Delta \rm z_{em}|$ and $|\Delta \rm M_i|$
reaching up to $>$1. The points for QSOs in S3 are shown in red.}
\end{figure}

\begin{figure}
\includegraphics[width=3.15in,height=4.0in]{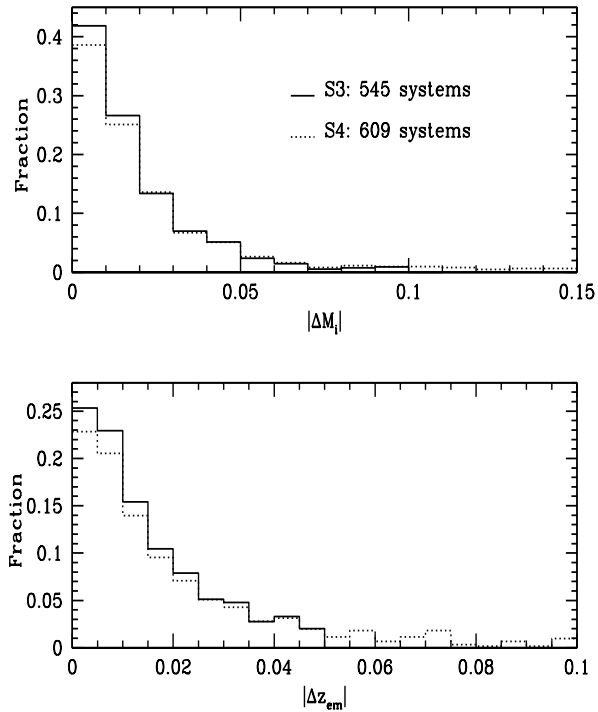}
\caption{Histograms showing absolute values of the difference between the i
magnitudes (upper panel) and emission redshifts (lower panel) of the
absorber-non-absorber QSO pairs for subsamples S3 and S4.}
\end{figure}

We use both types of comparison samples: one comprising of QSOs having no
absorbers along the line of sight (the non-absorber sample) and the other
comprising of QSOs having no DLAs along the line of sight irrespective of the
presence or absence of other absorption systems (the non-DLA sample). The
composite spectra of the non-absorber sample would be completely devoid of
reddening due to intervening absorbers and a comparison with the composite of
QSOs having DLAs along their lines of sight should reveal any weak reddening
present in these QSOs. On the other hand, reddening caused by any non-DLA
absorption systems would statistically cancel out when the composite of
absorber sample is compared with that of the matching non-DLA sample, giving an
estimate of reddening due to the DLAs alone. 

For compiling the matching non-absorber sample, we selected from DR7, QSOs
without any absorption systems of grade A and B. We chose to ignore any grade
C, D, E systems that may be present in the QSO spectra as these are not likely
to produce significant reddening as noted by Y06. Details of the system grades
can be found in Y06. In Fig.2 we have plotted the difference in M$_i$ and
z$_{em}$ between the full sample S1 and the corresponding non-absorber sample.
For some QSOs, differences are large, reaching up to $>$ 1. As identification
of absorption systems is difficult in the spectra of faint QSOs (which have
smaller S/N) we constructed a subsample of S1 having QSOs with Galactic
extinction corrected, i magnitude $<$ 19.5 so that the S/N of the corresponding
non-absorber spectra will be sufficiently high to ascertain the absence of
absorption systems. This subsample of S1 (S2) has 847 systems.  

The values of the differences in M$_i$ and z$_{em}$ of the pairs of QSOs in the
absorber subsample S2 and the corresponding non-absorber sample are also as
large as those for S1, and the match between M$_i$ and z$_{em}$ of the absorber
and non-absorber QSO pairs is not as good as that obtained for the lower
redshift Mg II sample of Y06 or the non-DLA-absorber sample of FP10.  For
better comparison of the composites, we therefore, chose two subsamples: S3
having 545 systems with $|\Delta$M$_i|<0.1$ and $|\Delta$z$_{em}| < 0.05$ and
S4 having 609 systems with $|\Delta$M$_i|<0.15$ and $|\Delta$z$_{em}| < 0.1$.
The redshift distributions for these subsamples are also shown in Fig.1.  About
94\% of the systems in these subsamples also have their redshifts between 2.25
and 4. The points corresponding to subsample S3 are shown in red in Fig.2.  The
histograms showing the distributions of $|\Delta$M$_i|$ and $|\Delta$z$_{em}|$
for these subsamples are shown in Fig.3. The values are mostly small and the
subsamples can be used to get reliable estimates of reddening. We constructed
the matching sample of non-DLAs in similar way.  The number of non-DLAs in SDSS
DR7 being very large, better matches were obtained between the parameters of
the QSOs in S1 and those of the matching non-DLA QSOs, with $|\Delta$z$_{em}|$
and  $|\Delta$M$_i|$ being smaller than 0.32 and 0.4 respectively.  
\section{Results}
\subsection{Reddening} The composites of S1 and of the corresponding
non-absorber sample, as well as the ratio of the two composites (the extinction
curve for S1) are plotted in Fig.4.  The best fit extinction curves for SMC as
well as for the MW are plotted.  Values of $E(B-V)$ are small.  $E(B-V)$ does
not change if we restrict the spectra to wavelengths smaller than 2400 {\AA}.
In Table 1 we give values of $E(B-V)$ with respect to both the non-absorber and
non-DLA matching samples for all subsamples, assuming the SMC extinction law.
Values of $E(B-V)$ for S3 and S4, using non-absorber samples, are 0.0044 and
0.0050 respectively. Formal errors are rather small and a bootstrap analysis
for S3 indicates 1 $\sigma$ error $\sim$ 0.0017 for both non-absorber as well
as non-DLA matching samples. Thus there is weak (2-3 $\sigma$) evidence for
reddening along the average DLA sight lines when compared to the non-absorber
sample. Using the non-DLA matching sample gives even smaller reddening (listed
in Table 1) which is expected as the reddening caused by other non-DLA
absorption systems along the line of sight could cancel out. 

Pettini et al. (1997) have argued that the dust to gas ratio is likely to be
smaller in high redshift DLAs as compared to the MW due to their lower metal
abundance, and lower depletion factors compared to that in the MW, resulting
in a small amount of extinction.  Following their argument, assuming the dust
to gas ratio to be 1/30$^{th}$ of that in the MW, and assuming N$_{\rm H
I}/E(B-V)\simeq 5\times 10^{21}$ cm$^{-2}$ for the MW (Diplas \& Savage 1994),
our subsamples S3 and S4 would have $E(B-V)$ of $\sim$ 0.003 in the absorber
restframe. This is considerably larger than,  but is within 2 $\sigma$ of, the
$E(B-V)$ obtained using the non-DLA matching sample.  It is somewhat smaller
than, but again within 1 $\sigma$ of, the value obtained using the non-absorber
matching sample.  However, we note that several of the QSOs in our DLA sample
should have lower redshift Mg II systems which should produce extinction as
found by Y06. Their full sample of 809 Mg II systems with W$_{Mg\,II}>0.3$
{\AA} has $E(B-V)$ of 0.01 which will correspond to an E(B-V) of 0.0053 in the
rest frame of the present DLA sample assuming the SMC extinction law. So the
$E(B-V)$ obtained using the non-absorber sample also seems to be much smaller
than the expected value, albeit within 3 $\sigma$ of it. Note that 0.0053
should roughly be the difference between the $E(B-V)$ values obtained using the
non-absorbers and non-DLA matching samples (values in columns 4 and 5 of table
1). This seems to be true for most of the subsamples having low reddening (for
which the difference of 0.0053 will be significant). So the method seems to be
giving consistent results and the average reddening in the whole DLA sample
appears to be very small. 

We do expect some DLAs at high redshifts to have higher reddening in view of
the following facts: (1) According to Noterdaeme et al.  (2008) at least 16\%
of the DLAs (having log(N$_{\rm H I})>20$) at high redshifts have molecular
hydrogen and there is a strong preference of H$_2$ bearing DLAs to have
significant depletion factors, 0.95$>$[X/Fe]$>$0.3, which could indicate
significant amount of dust in these systems. A correlation between molecular
hydrogen column density and colour excess is seen in the MW clouds (Rachford et
al. 2002) for $E(B-V)<0.1$.  (2) Some of the DLAs at high redshifts have been
found to have depletion patterns similar to those in the MW (Srianand et al.
2008), indicating the presence of dust similar to that in cold ISM of the
Galaxy. (3) DLAs at redshift $>$ 1.5 have been found to exhibit the 2175 {\AA}
bump (e.g. Noterdaeme et al. 2009a; Jiang et al. 2010) which again indicates
the presence of MW type dust in these absorbers. (4) A DLA at redshift of 2.45
in a GRB has the 2175 {\AA} bump in its rest frame (Eliasdottir et al. 2009). 
\begin{figure}
\includegraphics[width=3.15in,height=4.0in]{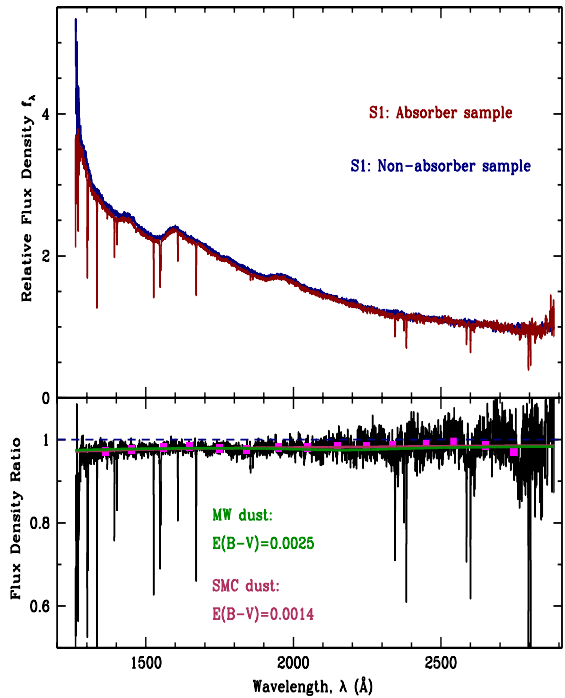}
\caption{The top panel shows the composite spectrum (geometric mean) of the
absorbers in subsample S1, in their rest-frames, in red, and that of the
corresponding non-absorbers, in blue. The two spectra have been normalized at
the higher wavelength end. The bottom panel shows the ratio of the two
spectra along with the best fit SMC extinction curve in red and MW extinction
curve in green. Pink squares are average flux density ratios over 100 {\AA}
bands, plotted to guide the eye.}
\end{figure}

\begin{figure*}
\includegraphics[width=5.00in,height=4.0in]{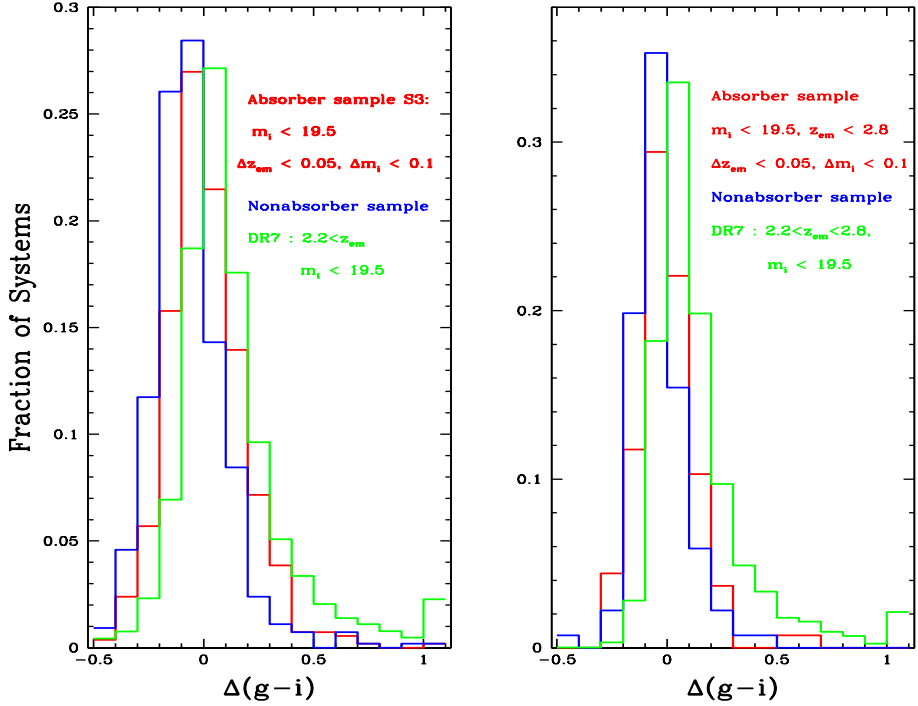}
\caption{(a; left panel) The figure shows the histogram, in red, of observer
frame colour excess $\Delta(g-i)$ for the subsample S3. The distribution for
corresponding non-absorber sample is shown in blue. The green histogram shows
the distribution for 9768 QSOs in DR7 having galactic extinction corrected i
magnitude smaller than 19.5 and having z$_{em}> 2.2$. (b; right panel) The
plots in this panel are similar except that only QSOs with $2.8>$z$_{em}> 2.2$
are included in all the three samples. There are 5378 DR7 QSOs in this plot.}
\end{figure*}

\begin{table*}
\centering
\begin{minipage}{140mm}
\caption{Mean and median values of $\Delta(g-i)$ for various subsamples}
\small
\begin{tabular}
{|l|l|l|r|r|r|r|l|l|l}
\hline
Sample  & \multicolumn{3}{c}{Absorber subsample}&\multicolumn{3}{c}{Non-absorber sample}& \multicolumn{3}{c}{DR7 subsample$^a$}\\
\hline
&Mean&Sigma&median&Mean&Sigma&median&Mean&Sigma&median\\
\hline
S3&0.02&0.20&-0.001&-0.058&0.19&-0.075&0.14&0.31&0.075\\
S4&0.02&0.22&0.000&-0.058&0.20&-0.076&&&\\
S3$^b$&0.00&0.14&-0.009&-0.032&0.12&-0.042&0.16&0.29&0.085\\ 
\hline
\multicolumn{10}{l}{ a: DR7 QSOs having m$_i<19.5$ and 2.2$<$ z$_{em} < 5.2$ }\\
\multicolumn{10}{l}{ b: Values in this row are for systems with z$_{em} < 2.8$}\\
\end{tabular}
\end{minipage}
\end{table*}
\begin{table*}
\centering
\begin{minipage}{140mm}
\caption{Relative reddening between subsamples}
\small
\begin{tabular}
{|r|r|r|r|r|r|r|}
\hline
\multicolumn{2}{c}{Sample A}&\multicolumn{2}{c}{Sample
B}&\multicolumn{3}{c}{$E(B-V)^a$}\\
Subsample&Criterion&Subsample&Criterion&SMC&LMC&MW\\
\hline
S5&log(N$_{\rm HI}) \ge 20.49$&S6&log(N$_{\rm HI)} <
20.49$&0.000&-0.002&-0.003\\
S7&m$_i\ge$19.05&S8&m$_i<$19.05&0.004&0.025&0.044\\
S9&M$_i\ge$-27.42&S10&M$_i<$-27.42&-0.006&0.001&0.020\\
S11&z$_{ab}\ge2.89$&S12&z$_{ab}<2.89$&-0.000&-0.024&-0.036\\
S13&z$_{em}\ge3.26$&S14&z$_{em}<3.26$&0.009&0.020&0.018\\
S15&$\Delta(g-i)\ge-0.01$&S16&$\Delta(g-i)<-0.01$&0.018&0.031&-0.006\\
S17&W$_{\rm Si II}\ge0.62$&S18&W$_{\rm Si II}<0.62$&0.006&0.021&0.015\\
S19&log(N$_{\rm HI})\ge21.0$&S20&log(N$_{\rm HI)}< 21.0$&0.006&-0.001&-0.017\\
S21&$\Delta(g-i)\ge0.30$&S22&$\Delta(g-i)<0.30$&0.038&0.069&0.010\\
S23&W$_{\rm Si II}\ge1.0$&S24&W$_{\rm Si II}<1.0$&0.008&0.019&0.006\\
S25&W$_{\rm Si II}\ge1.5$&S26&W$_{\rm Si II}<1.5$&0.006&0.015&0.009\\
S22&$\Delta(g-i)<0.30$&S16&$\Delta(g-i)<-0.01$&0.007&0.013&0.000\\
\hline
\multicolumn{5}{l}{$^a$ Relative extinction of subsample A with respect to subsample
B }\\
\end{tabular}
\end{minipage}
\end{table*}

\subsubsection{Distribution of $\Delta(g-i)$} To explore the issue of dust in
DLAs further, we looked at the distributions of the $\Delta(g-i)$ values of the
absorber and the non-absorber sample. These are shown in Fig.5a for subsample
S3. The distributions do appear to be different.  The mean, sigma and median
values for the absorber and the non-absorber samples are given in Table 2.
Similar values are also given for S4. The 1 $\sigma$ error on the difference in
the mean values of $\Delta(g-i)$ is $\sim$ 0.012.Thus the $\Delta(g-i)$ values
of the absorber and non-absorber samples differ at 6 $\sigma$ level. The
Kolmogorov-Smirnov (KS) probabilities for the distributions of $\Delta(g-i)$
for the absorber and non-absorber samples to be similar is close to zero for
both S3 and S4. We note that the KS test compares the shapes of distributions
of the $\Delta(g-i)$ values of the absorber and the non-absorber sample rather
than comparing their average or median values, and gives the probability for
the two distributions being drawn from a parent distribution. We have also
plotted in Fig.5a the  $\Delta(g-i)$ distribution for the 9768 DR7 QSOs having
z$_{em} > 2.2$ and m$_i < 19.5$.  Values of the mean and variance for this
sample are also included in Table 2.  The mean values are different at high
significance level from those for the absorber as well as the non-absorber
sample. The KS probability for the distribution of $\Delta(g-i)$ in DR7 plotted
in the figure and the absorber subsamples S3 and S4 as well as corresponding
non-absorber samples to be similar is also close to 0. This could partly be due
to the presence of BALQSOs in the DR7 sample as seen from the high
$\Delta(g-i)$ tail of the green histogram in Fig.5a and Fig.5b. 

The KS test shows that the $\Delta(g-i)$ distributions of the absorber and
non-absorbers samples are different. Those distributions are also different
than the distributions of intermediate redshift Mg II systems described by Y06;
there is a clear difference between Fig.5a and the Fig.3 of Y06. In the former,
the red tail towards the high $\Delta(g-i)$ is clearly missing. It may,
however, be noted that the $(g-i)$ colour loses its meaning for QSOs with
z$_{em} > 2.8$ as for these QSOs, the $g$ band falls in Lyman alpha forest and
can be dictated by fluctuations from object to object in the total attenuation
in the Lyman alpha forest. Only 136 out Of the 847 QSOs in S2 have emission
redshifts smaller than 2.8. Thus comparison between $\Delta(g-i)$ values for
the absorber and non-absorber samples may not be very meaningful as indicators
of dust content of the absorbers. In Fig.5b we have plotted $\Delta(g-i)$
distributions for the three samples as in Fig.5a, restricting to QSOs having
z$_{em}<2.8$.  The statistical details for these samples are included in the
last row of Table 2.  The mean and median values for the absorber and
non-absorber samples now differ at 2 $\sigma$ level and the KS probability for
the distributions of $\Delta(g-i)$ for the absorber and non-absorber samples to
be similar is 0.07 indicating smaller difference in the distributions of
spectral shapes for the absorber and non-absorber samples. The KS
probabilities for the $\Delta(g-i)$ distributions of the 5378 DR7 QSOs (having
z$_{em}$ between 2.2 and 2.8 and m$_i<19.5$) and those of the absorber and
non-absorber samples, plotted in Fig.5b, to be similar are still close to
zero.

\begin{figure}
\includegraphics[width=3.00in,height=3.5in]{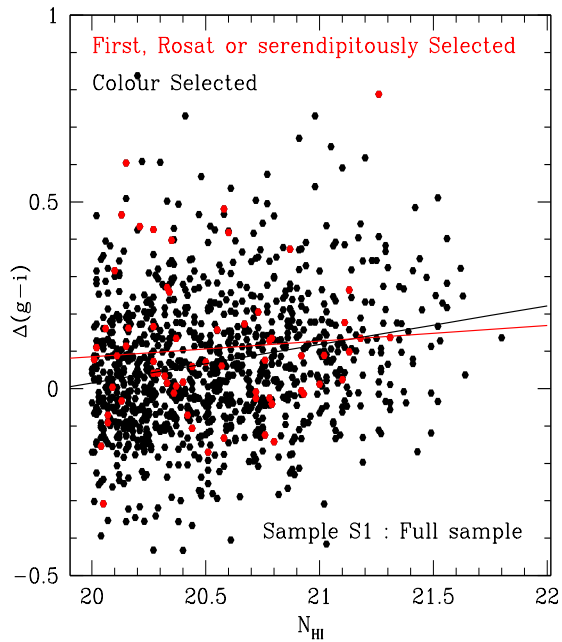}
\caption{Plot showing observer frame colour excess $\Delta(g-i)$ as a function
of H I column density in individual DLAs in S1. QSOs not selected on the basis
of colours but selected on the basis of having been detected in FIRST or ROSAT
survey or detected serendipitously are shown in red. The weak correlation for
the full sample (slope of best fit line = 0.103) is possibly due to the fact
that systems having higher H I may increase the g-magnitudes more due to their
stronger Lyman $\alpha$ lines which lie in the g band for QSOs with
z$_{em}>2.8$. These QSOs form 87\% of sample S1. For the remaining 13\% of QSOs
with z$_{em}\le2.8$ the best fit line is flatter having a slope of 0.027. Red line is the bestfit line for the red points and is flatter (slope 0.042) than the black line.}
\end{figure}

In Fig.6, we have plotted $\Delta(g-i)$ vs. N$_{\rm HI}$ for the entire sample
S1. There does exist a weak correlation between the two quantities which could
result due to the fact that stronger Lyman alpha lines, which lie in the g band
for most (84\%) of the QSOs (with z$_{em}>2.8$), will result in higher g
magnitudes. The slope of the best fit line drawn in the figure, is 0.103.
Restricting to the 136 QSOs with z$_{em}<2.8$, the correlation is much weaker,
the slope of the best fit line being 0.027. This lends support to the above
argument. In this figure we have indicated the QSOs which have not been colour
selected but have been selected based on their detection in the radio or X-ray
bands or have been discovered serendipitously. The best fit line for these is
flatter and has a slope of 0.042. The slope of bestfit line for purely colour
selected QSOs (0.107) is very close to that for the full sample S1. The KS test
shows that the probability that the $\Delta(g-i)$ distribution for the colour
selected and rest of the QSOs in S1 to be same is 0.2. 

\subsubsection{Dependence on QSO and absorber properties} To explore the
issue of reddening further we tried to study the dependence of the slope of the
composite spectra on the absorber  and QSO properties. We divided the absorber
sample in two roughly equal halves (see Table 1) depending on N$_{\rm HI}$,
z$_{em}$, z$_{ab}$, m$_i$, M$_i$, $\Delta(g-i)$ and W$_{\rm Si II}$ (rest
equivalent width of the Si II $\lambda1526$) (subsamples S5-S18).  Over the
redshift range covered by our DLA sample, Mg II doublets fall outside the SDSS
spectra in most cases and the most convenient lines to use as surrogates of Mg
II selection are Si II $\lambda1526$ and Al II $\lambda1670$.  Even though
these lines are expected to be weaker than the Mg II doublet lines, they are
still near saturation. We also constructed a few subsamples with different
limiting values of N$_{\rm HI}$, $\Delta(g-i)$, and  W$_{\rm SiII}$ (subsamples
S19-S26). The $E(B-V)$ values in Table 1 show that the reddening is higher for
subsamples with (i) higher  N$_{\rm HI}$, (ii) fainter QSOs, (iii) lower
z$_{ab}$, (iv) higher W$_{\rm Si II}$ and (v) higher $\Delta(g-i)$. We have
also confirmed that the conclusions for subsamples based on the equivalent
widths of Si II$\lambda1526$ also hold for subsamples based on the equivalent
widths of Al II$\lambda1670$. 

In Fig.7, we have plotted the composites of five sets of subsamples together.
The spectra are normalized at the long wavelength end. It can be seen that the
slope of the composite indeed depends on the absorber properties. Subsamples
based on W$_{\rm SiII}$ and $\Delta(g-i)$ show definite relative reddening.  We
also constructed relative extinction curves for pairs of subsamples by dividing
the composite spectrum of one subsample with that of another. Best fit values
of $E(B-V)$ for all three types of extinction curves, SMC, LMC and MW, are
given in Table 3. The three values for given pairs of subsamples are different.
The quality of fit in the three cases is similar, SMC giving the smallest
$\chi^2$ values in most cases.  Most of the $E(B-V)$ values are small. We
conclude that though the reddening is not large, there is definite signature of
reddening for subsamples divided with respect to  N$_{\rm HI}$, W$_{\rm SiII}$
and $\Delta(g-i)$. We further discuss this below.
\subsubsection{Dependence on the mode of selection of SDSS QSOs} We studied the
dependence of the dust content of the DLAs on the mode of selection of SDSS QSO
targets. It has been argued that dusty QSOs might be missed due to the process
of colour selection of targets in SDSS QSO survey (M{\'e}nard et al. 2008). In
sample S1, 65 QSOs have not been selected on the basis of their colours but
have been selected on the basis of their radio or X-ray properties or have been
serendipitously detected.  We have constructed composite of this sample (NCS65
sample) and also of the sample of the rest of the 1019 colour selected QSOs
(CS1019 sample) in sample S1. The relative $E(B-V)$ of the NCS65 with respect
to CS1019 is 0.0027. The emission redshift distributions of these two samples
are however different, KS test giving the probability that the two sets of
redshifts to be drawn from the same distribution close to zero, the CS1019
sample having higher redshifts. Thus the relative reddening could partly be due
to the difference in emission redshifts of the two samples (e.g. see samples 13
and 14 in tables 1 and 3).  We therefore selected QSOs from the colour selected
sample of 1019 systems (CS1019) which have z$_{em}$ and m$_i$ values close to
those of the QSOs in the sample which were not colour selected (NCS65), using
the procedure described in section 2.  This new sample of 65 colour selected
QSOs (CS65) has similar distribution of z$_{em}$ and m$_i$ as that for the
sample of QSOs which are not colour selected (NCS65), KS test yielding the
probability for the distributions to be same to be $>$ 0.93. The relative
$E(B-V)$ of NCS65 with respect to CS65 is 0.0020. So the QSOs selected by
methods other than colour selection appear to be more reddened as compared to
the colour selected QSOs. We have plotted composites for CS65 and NCS65 in the
bottom panel of Fig.7. We note that only 4 systems are common in NCS65 and S23.

\subsection{Line strengths}
We next studied the absorption lines in the composite spectra in order to
examine any correlation with absorber and QSO properties. The arithmetic mean
composite of the sample S1 is shown in Fig.8. The equivalent widths of
absorption lines for this spectrum are given in Table 4. The table also
includes equivalent widths of the sample 1 (full sample) of Y06.  The DLA
systems appear to have lower ionization compared to the full sample of Y06 as
seen from the considerably lower equivalent widths of C IV and Al III lines.  

In order to study the dependence of line strengths on the absorber and QSO
properties we have plotted in Fig.9 lines of several species for some of
the subsamples from Table 1. Lines for two subsamples (in red and in black) are
plotted in each row; the details of the subsamples are given in the last column
of that row. Also plotted in green, are the 1 $\sigma$ errors in flux of the
composite spectra. As expected, lines of higher ionization are somewhat weaker
and those of lower ionization are somewhat stronger in the subsample having
higher H I column density. Subsample S11 with higher z$_{ab}$ seems to have
somewhat weaker lines of higher ionization species as compared to the
corresponding subsample S12 at lower redshift. The difference is not very
significant for Si IV lines. The difference is not likely to be due to the
slight difference in the average H I column densities in these subsamples. If
true this may indicate a change in the ionizing radiation.

Line strengths do seem to depend on the limiting values of $\Delta(g-i)$
and more strongly on W$_{\rm SiII}$ as seen from the 5$^{th}$ and 6$^{th}$ rows
in Fig.9. In the third from bottom panel of the figure we have compared lines
for subsamples S16 and S22.  The equivalent widths of lines of subsamples
having $\Delta(g-i)<-0.01$ (S16) and $\Delta(g-i)<0.3$ (S22) are very similar.
The relative $E(B-V)$ between these two subsamples is 0.007 (see Table 3).
This indicates that the systems with $\Delta(g-i)<0.3$ are basically all
similar and it is only a few systems with $\Delta(g-i)\ge0.3$ that have larger
equivalent widths and also higher H I column densities (see Table 1). Only 13
systems are common between samples S21 and S23. Removing these systems from S21
makes the line strengths the same as those for S22 (indicating that the higher
$\Delta(g-i)$ in rest of the systems in S21 may not originate in the
intervening absorbers and could be intrinsic to the quasar), while removing
these systems from S23 still leaves lines significantly stronger than those for
S24. Thus we conclude that the line strengths are most sensitive to W$_{\rm
SiII}$. S23 forms $\sim$ 10\% of the DLA sample. Some of these could be the
dusty, high abundance systems found at high redshifts as mentioned in section
3.1. In the second from bottom panel of the figure we have compared lines for subsamples S19 and S23. Lines of S23 seem to be stronger than those for S19.

In the bottom panel of Fig.9, we have plotted lines for the sample of QSOs
based on colour selection and not based on colour selection (NCS65 and CS65 as
described in section 3.1.3). Lines of C and Al are weaker in the colour
selected sample, those of Si are of similar strength, while all lines of Fe II
(only one of which is included in Fig.9) are stronger in this sample. The lines
O I$\lambda1302$ and Si II$\lambda1304$ are exceptionally weak in CS65 which
can not be real in view of the strength of Si II$\lambda1526$ line. We believe
this may be caused by the fact that the samples are small and for almost half
of the samples these lines lie in the Lyman alpha forest  and thus may be
severely affected. The average N$_{\rm HI}$ of both samples differ by only
$\sim$ 0.1 dex. The composite spectra thus may indicate lower abundances in the
colour selected sample and higher depletion of Fe and possibly Si in the sample
not selected on the basis of colour.  This is consistent with higher dust
content of the DLAs along lines of sight to QSOs which are not colour
selected.

\begin{table}
\centering
\begin{minipage}{100mm}
\caption{Equivalent widths in m{\AA} for metal absorption features\hfill\break
in the absorber rest frame in sample S1} 
\begin{tabular}
{|l|l|r|r|}
\hline 
&&\multicolumn{2}{c}{Sample}\\
Wavelength&Species&S1&1(Y06)\\
\hline
1302.17&O I&413$\pm$14&\\
1304.37&Si II&266$\pm$14 &\\
1334.53&C II& 575$\pm$9&\\
1370.13&Ni II&47$\pm$5&\\
1393.32&Si IV&300$\pm$7&\\
1402.77&Si IV&230$\pm$7&\\
1526.71&Si II&447$\pm$4&383$\pm$10\\
1548.20&C IV&511$\pm$5&839$\pm10$\\
1550.78&C IV&300$\pm$21&580$\pm9$\\
1608.45&Fe II&227$\pm$4&\\
1656.93&C I&$<$12&19$\pm4$\\
1670.79&Al II&458$\pm$4&471$\pm6$\\
1709.60&Ni II& 49$\pm$5&22$\pm$5\\
1741.55&Ni II&16$\pm$5&28$\pm$4\\
1751.92&Ni II&28$\pm$4&$<11$\\
1808.00&Si II&51$\pm$2&68$\pm5$\\
1827.94&Mg I&$<7$&$<$5\\
1854.72&Al III&107$\pm$3&183$\pm4$\\
1862.79&Al III&47$\pm$2&112$\pm4$\\
2026.48&Mg I+Zn II&33$\pm$5&40$\pm$4\\
2056.26&Cr II&26$\pm$3&39$\pm4$\\
2062.24&Cr II+Zn II&27$\pm$4&48$\pm$4\\
2066.16&Cr II&$<8$&14$\pm4$\\
2249.88&Fe II&32$\pm$6&35$\pm$4\\
2260.78&Fe II&38$\pm$4&46$\pm$4\\
2344.21&Fe II&462$\pm$8&550$\pm4$\\
2367.59&Fe II&$<15$&$<4$\\
2374.46&Fe II&245$\pm$11&298$\pm4$\\
2382.77&Fe II&753$\pm$9&761$\pm4$\\
2576.88&Mn II&36$\pm$8&54$\pm4$\\
2586.65&Fe II&491$\pm$16&509$\pm4$\\
2594.50&Mn II&$<27$&50$\pm4$\\
2600.17&Fe II&724$\pm$13&783$\pm4$\\
2606.46&Mn II&$<16$&26$\pm4$\\
2796.35&Mg II$^a$&1698$\pm$80&1432$\pm4$\\
2803.53&Mg II&836$\pm$73&1261$\pm4$\\
\hline
\multicolumn{4}{l}{ a: Only 22 systems have Mg II inside SDSS spectra}\\
\end{tabular}
\end{minipage}
\end{table}

\begin{figure}
\includegraphics[width=3.00in,height=5.5in]{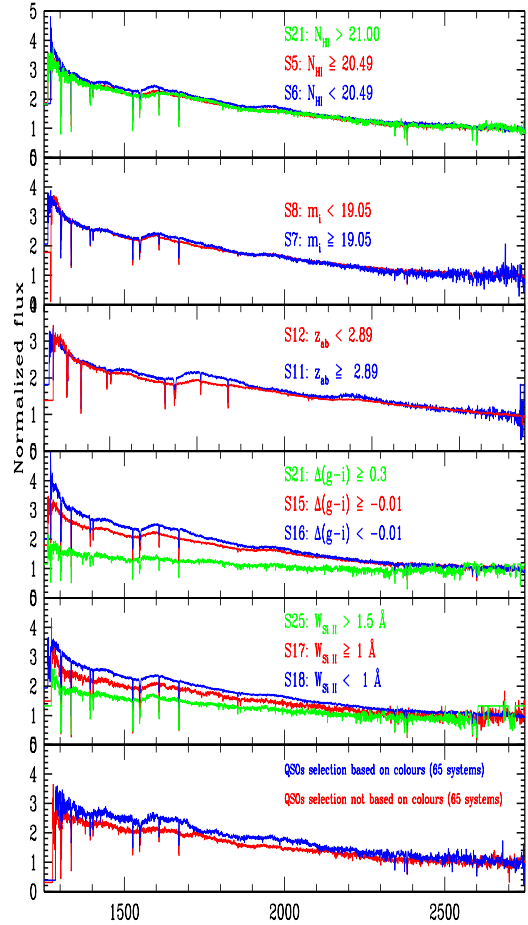}
\caption{The geometric mean composite spectra of five sets of subsamples are
plotted in the top five panels.  The spectra are normalized at the long
wavelength end. The subsamples are detailed in the legends. The scales along
the x axis are the same for all panels except for the third panel where it goes
from 1250-2350 {\AA} as the low redshift composite spectra does not extend
beyond that. The dust content is most clearly correlated with Si
II$\lambda1526$ equivalent width and $\Delta(g-i)$. In the bottom panel we have
plotted the composites for DLAs along the lines of sight to colour selected
QSOs (CS65) and along the lines of sight to QSOs not selected on the basis of
colour (NCS65). The latter sample is more reddened compared to the colour
selected sample. Note that both samples match in emission redshift and i
magnitude distribution.}
\end{figure}
\begin{figure*}
\includegraphics[width=7.50in,height=6.0in]{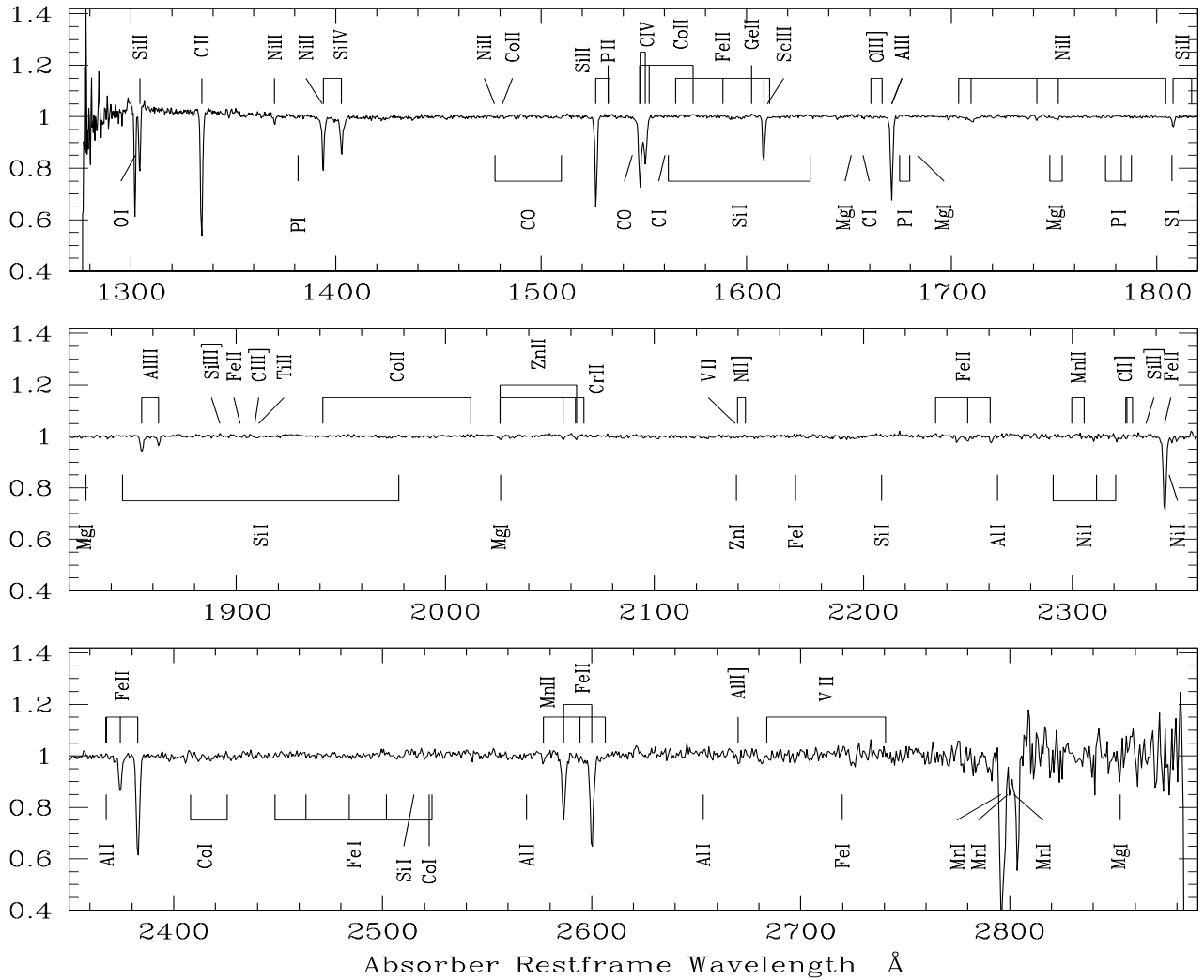}
\caption{The arithmetic mean composite spectrum of the systems in the full
sample (S1) in the absorber rest frame. The equivalent widths of the absorption
lines are given in Table 4.}
\end{figure*}

%\begin{figure*}
%\includegraphics[width=7.50in,height=8.0in]{F9.jpeg}
\noindent {\bf Fig.9 (This figure can be seen as jpeg image) Caption:}
Comparison of several absorption lines in the composite spectra of a number of
subsamples, defined in Table 1, constructed on the basis of absorber and QSO
properties.  The lines for the full sample are plotted in the top panel. Each
row shows different lines in the composite spectra of the same two subsamples
(one in red and the other in black), which are detailed in the last column of
each row.  Each column contains lines of the same species, indicated in the top
row of that column, the wavelength of which is indicated below each column. One
$\sigma$ errors in the flux of the composite spectra are shown in green for
each pixel.  Scales are identical for all rows of a given column, except for
the second from bottom panel, for which y axis scale goes to smaller values
because of the stronger lines for the subsamples plotted in that row.  The tick
marks on the y axis are 0.1 units apart. The x axis spans 10 {\AA} in all
columns. 
%\end{figure*}

\subsection{Abundances}
For the weakest lines in the composite spectrum, the least affected by
saturation, the equivalent widths in the full Y06 sample and those of S1 are
nearly the same (see Table 4), implying the typical column densities are
similar (to within factors of 2) while the average column density of N$_{\rm H
I}$ differs by a factor of 5 (log $<$N$_{\rm H I}>$ = 20.72 for S1 and equals
20.00 for the full sample of Y06).  Evidently, all of the involved species (Si
II, Zn II, Cr II, Fe II, Mn II) are more deficient compared to H I in the
sample S1 of this paper.  The composite spectrum thus implies lower abundance
of heavy elements, in general, at high redshifts than at lower redshifts in the
regions probed by QSO absorption systems. In Table 5 we have given the
abundances obtained from weakest lines of the species mentioned above, assuming
them to be on the linear part of the curve of growth.  The abundances are about
4\% of the solar values, with Fe and Mn having somewhat lower abundances. The
abundances are consistent with the predictions of chemical evolution model of
Pei et al. (1999). We however, note that these values are obtained from single
lines and may not be robust. 
\begin{table}
\centering
\begin{minipage}{100mm}
\caption{Abundances as obtained from weakest lines}
\begin{tabular}
{|c|c|c|c|c|c|c|}
\hline 
Sample&[Si/H]&[Cr/H]&[Mn/H]&[Fe/H]&[Ni/H]&[Zn/H]\\
S1&-1.29&-1.49&-1.99&-1.63&-1.35&-1.38\\
\hline
\end{tabular}
\end{minipage}
\end{table}

As noted above, the composite spectrum for subsample S23 shows stronger lines
compared to those of all other subsamples, including subsample S5 (log(N$_{\rm
HI})\ge20.48$) which has the same average H I column density and compared to
those of subsample S19 which has almost twice the H I column density of as that
for S23. This is also true for few other weak lines which are seen in the
spectrum, namely, Al III$\lambda\lambda$ 1862 and Ni II $\lambda$1741. Thus it
appears that the average abundance of S23 is higher than the rest of the
sample.  As noted above, the abundances of elements other than Fe seem to be
lower in DLAs along the lines of sight of colour selected QSOs.

\section{Conclusions} We have used the method of composite spectra to
investigate the dust content of DLAs in the SDSS DR7 QSOs. We find
definite evidence of dust, the reddening depending most sensitively on the
equivalent width of Si II$\lambda1526$ and Al II$\lambda1670$ lines and as
expected the $\Delta(g-i)$ values. The spectra of 80\% of QSOs having high
($>$0.3) values of $\Delta(g-i)$ appear to be reddened intrinsically to the
QSOs.  The average differential extinction in DLA systems appears to be very
small. We also compared the strengths of absorption lines in the composite
spectra of different subsamples. The lines are strongest for the subsample with
values of W$_{\rm Si II}>1.0$. Based on these studies, we conclude that
$\le$10\% of the DLAs in our sample have significant amount of dust, and have
higher abundances as compared to the rest of the sample. 

The reason for the difference between the dust content of {Mg II} samples
investigated by Y06 and the DLA sample studied in this paper could possibly be
the difference in the Mg II equivalent widths and redshifts of the two samples.
Mg II $\lambda2796$ is covered in only 22 of the SDSS spectra of the DLA
sample, so we are unable to get meaningful value of the average equivalent
width. The average redshift of S3 is 2.84 (Table 1). From Table 1 of Y06 we
see that their full sample, sample 1, has an average redshift of 1.33 and
$E(B-V)$ of 0.01. Assuming that the redshift dependence of absorber restframe
$E(B-V)$ derived by M{\'e}nard et al.  (2008) namely, $E(B-V)\propto (1+{\rm
z}_{ab})^{-1.1}$ for z$_{ab}<2$ is valid at higher redshifts, sample S3
(assuming it has similar average Mg II equivalent width) can be expected to
have $E(B-V)$ of $\sim$ 0.006 which is higher than the $E(B-V)$ obtained for
the full sample. 

As found here, most DLAs do not have significant amount of dust while 10\%
of the observed DLAs do have some dust and likely higher abundances. Some of
these could be the dusty systems found by Srianand et al. (2008). It is
possible that there is a bimodal distribution of dust in DLAs as was pointed
out by Khare et al. (2007): there may be systems with much higher dust content
and QSOs behind such dusty systems are too faint to be seen in magnitude
limited surveys and/or are  too reddened to be included in the colour selected
samples.  Even though no evidence for this is found from studies of radio
selected samples of QSOs (e.g. Ellison et al. 2001; Akerman et al. 2005;
Jorgenson et al. 2006), it may be worth noting that a few of such QSOs have
been observed by SDSS through other means of selection (see e.g.  Noterdaeme et
al.2009a) and have higher dust content and higher abundances. The SDSS data
base would thus have very few of such systems.  M{\'e}nard et al. (2008) have
shown that the fraction of missed absorbers rises from 1 to 50\% with increase
in Mg II rest equivalent width from 1 to 6 {\AA}.

We find a significantly higher dust content in DLAs along the lines of
sight to SDSS QSOs that have not been selected on the basis of colours. Though
the difference in the dust content of the two samples split by method of
selection is not large, the finding is consistent with the expectations that
dusty QSOs might be missed in SDSS due to colour selection bias. Therefore, the
fraction of DLAs we find containing dust in the SDSS ($\sim$10\%) may be
considered a lower-limit with respect to the global sample.

A larger survey of high-z QSOs selected without regard to optical colors is
needed to more accurately measure the fraction of dusty DLAs at these
redshifts.  A substantial fraction of the z$>$2.2 quasars now being observed by
the SDSS-III Baryon Oscillation Spectroscopic Survey have been targeted by
alternate methods (Ross et al. 2011).  These spectra may provide the necessary
statistics to better explore the dust content of DLAs at high-z.  

Should the small fraction of dusty DLAs hold in a less-biased sample, one might
speculate that the dust grain sizes may be different (larger) at high redshifts
than those at lower redshifts.  If the grains are large, there is no selective
extinction at wavelengths below about 1/6 the smallest grain size in a mix
which can explain the absence of reddening. The distribution of grain sizes and
the extinction curve may then be changing with cosmic time.  Examples of such
flat extinction curves exist in the Galaxy, when R$_V$ = 5 instead of the
normal value of three (e.g. Mathis 1990).  We reserve these conclusions for
future work.

\section*{Acknowledgments}
PK thanks the Council of Scientific and Industrial Research, India for
financial support. 

Funding for the creation and distribution of the SDSS Archive has 
been provided by the Alfred P. Sloan Foundation, the Participating 
Institutions, the National Aeronautics and Space Administration, the 
National Science Foundation, the U.S. Department of Energy, the 
Japanese Monbukagakusho, and the Max Planck Society. The SDSS Web 
site is http://www.sdss.org/.

The SDSS is managed by the Astrophysical Research Consortium (ARC) 
for the Participating Institutions. The Participating Institutions 
are The University of Chicago, Fermilab, the Institute for Advanced 
Study, the Japan Participation Group, The Johns Hopkins University, 
the Korean Scientist Group, Los Alamos National Laboratory, the 
Max-Planck-Institute for Astronomy (MPIA), the Max-Planck-Institute 
for Astrophysics (MPA), New Mexico State University, University of 
Pittsburgh, University of Portsmouth, Princeton University, the 
United States Naval Observatory, and the University of Washington.\\

\end{document}